# Hidden Division of Labor in Scientific Teams Revealed Through 1.6 Million LaTeX Files

**Authors:** Jiaxin Pei[1,2*], Lulin Yang[1], Lingfei Wu[1*]

**Affiliations:**

[1] School of Computing and Information, The University of Pittsburgh, 135 N Bellefield Ave, Pittsburgh, PA 15213

[2] Stanford Institute for Human-Centered Artificial Intelligence, 353 Jane Stanford Way, Stanford, CA 94305

[*]Corresponding author. E-mail: liw105@pitt.edu (L. W.) and pedropei@stanford.edu (J. P.)

## Abstract

Recognition of individual contributions is fundamental to the scientific reward system, yet coauthored papers obscure who did what. Traditional proxies—author order and career stage—reinforce biases, while contribution statements remain self-reported and limited to select journals. We construct the first large-scale dataset on writing contributions by analyzing author-specific macros in LaTeX files from 1.6 million papers (1991–2023) by 2 million scientists. Validation against self-reported statements (precision = 0.87), author order patterns, field-specific norms, and Overleaf records (Spearman's $\rho = 0.6$, $p < 0.05$) confirms the reliability of the created data. Using explicit section information, we reveal a hidden division of labor within scientific teams: some authors primarily contribute to conceptual sections (e.g., Introduction and Discussion), while others focus on technical sections (e.g., Methods and Experiments). These findings provide the first large-scale evidence of implicit labor division in scientific teams, challenging conventional authorship practices and informing institutional policies on credit allocation.

## Main

Scientific progress relies on a reward system that recognizes contributions, but this system is imperfect. In the 1960s, Robert Merton observed that citations—used as a proxy for recognition—disproportionately favored established scientists over equally deserving junior peers, a phenomenon he termed the *Matthew Effect* [1]. He explained this by selective memory: crediting a few scientists for many discoveries simplifies recall and communication. However, this also means most scientists' contributions go unrecognized.

The gap between individual contribution and recognition will likely widen in coauthored work, as citations credit papers collectively rather than individually, further obscuring who did what in the team. This issue has intensified as research shifts from solo efforts to teamwork: solo authorship declined from nearly 80% in the 1960s to 50% in sociology, 26% in economics, and just 7% in computer science by the 2010s[2–4]. Meanwhile, large-scale collaborations are now essential to advancing research. For example, mapping the human genome (2001) engaged more than 2,800 researchers from six countries[5], while discovering gravitational waves (2016) involved over 3,500 authors across 18 countries[6]. In teams of this scale, determining who does what is nearly impossible—let alone ensuring proper credit.

To keep up with the rise of scientific teams, authorship practices have evolved beyond author order, with self-reported contribution statements emerging to distinguish and acknowledge authors' roles within teams. The biomedical and life sciences fields were among the first to adopt this practice, as lab-based research is the conventional mode of scientific inquiry in these disciplines. In the early 2000s, journals such as *Nature* and *PNAS* formally recognized this approach by allowing contribution statements in published articles[7]. Following this, disciplines like psychology, physics, and chemistry increasingly adopted contribution statements, reflecting the rise of interdisciplinary and large-team collaborations. By 2010, major publishers such as Elsevier, Springer, and Wiley had begun incorporating author contributions into their policies, though widespread adoption accelerated in the following decade. By 2020, standardized crediting frameworks, such as CRediT (Contributor Roles Taxonomy), had been adopted by more publishers, ensuring greater accountability and recognition in collaborative research[8].

While these developments in contribution tracking go beyond traditional proxies that rely on author order or career stage—both of which reinforce social biases—they still have several limitations. First, they are self-reported and often determined by the corresponding or senior author, introducing potential biases. Second, they rely on predetermined roles, such as those defined by the CRediT system, which may not accommodate rapidly evolving roles and new forms of collaboration. Finally, they are limited to select journals, restricting their broader applicability. Therefore, despite the emergence of self-reported contributions and previous studies[9–13], a fundamental question remains: How do scientific teams divide work?

Theoretical perspectives suggest that innovation in science and technology involves two key dimensions—idea generation and idea execution[9–11,14]. Idea generation involves formulating research questions, developing hypotheses, and constructing analytical frameworks, whereas idea execution entails conducting data analysis, implementing experiments, and synthesizing findings into written reports. Inspired by this insight, and in contrast to the common belief that author contributions simply decline with author order, we hypothesize that this pattern depends on the nature of the work. Specifically, we argue that scientific teams exhibit a systematic division of labor between conceptual and technical work, with author contributions distributed unevenly across article sections.

Testing this hypothesis is challenging, as it requires identifying authors' actual contributions beyond author ranks or self-reports—an area rarely explored. To address this gap, we developed a novel method to quantify author writing contributions using author-specific macros in LaTeX source code files, inspired by prior work analyzing scientific writing conventions[15]. We collected LaTeX files underlying 1.6 million arXiv papers by 2 million scientists. Established in the 1990s, arXiv.org is the largest preprint repository for STEM fields, including mathematics, statistics, computer science, and more, where LaTeX—a typesetting system widely used for formatting equations—is commonly adopted. Using this dataset, we identified contributions by tracking unique macros: if an author previously used a macro appearing in a paper, they were considered a potential contributor. Macros were attributed to multiple authors if they matched their individual records. Contribution share was estimated by normalizing unique macros per author across a paper. We successfully identified the contributions of 583,817 scientists across 730,914 papers (1991-2023).

We validate this dataset using four complementary approaches. First, we compared it against 469 self-reported author contributions from four journals—*Science, Nature, PNAS,* and *PLOS ONE*—confirming high precision (0.87) and recall (0.71) in identifying paper-writing contributions. Second, we analyzed the relationship between author order and contribution, confirming that contributions decrease with author order, consistent with prior assumptions[16].

Third, we examined field-specific patterns, showing that our estimated contributions align with disciplinary norms, such as alphabetical ordering in economics and first-author emphasis in computer science[17]. Finally, we validated our approach using 14 detailed contribution records from Overleaf, finding a Spearman's ρ = 0.6 ($p < 0.05$) between contribution ranks estimated from LaTeX code analysis and actual editing activity.

After confirming data quality, we leveraged explicit section information in LatTeX files to examine the distribution of work across sections. Our analysis reveals a hidden division of labor: some authors primarily contribute to conceptual sections (e.g., Introduction, Discussion), while others focus on technical sections (e.g., Methods, Experiments). Though long assumed, this study provides the first large-scale evidence of implicit labor division in scientific teams. These findings challenge conventional authorship practices and have implications for credit allocation in hiring, promotion, and funding decisions.

**Results**

**Hidden Division of Conceptual and Technical Labor in Scientific Teams**

Scientific research is carried out by teams in which members contribute in distinct ways based on experience, expertise, and seniority. Senior authors, typically listed later, shape the project's theoretical framework and analysis, while junior authors, listed earlier, focus on data collection, preliminary analysis, and drafting. By analyzing LaTeX source files from 411,808 (1994-2020) papers, we systematically uncover a structured division of labor between conceptual and technical work.

Fig. 1 illustrates section-specific patterns in author order. By extracting major sections (e.g., Introduction, Methods, Results) using \section tags in LaTeX files, we normalize the number of unique macros within each section relative to all sections for a given author as a proxy for their "attention" or focus of work. We then build regression models to predict contribution based on author rank, obtaining statistically significant estimates of these effects. The results show that the first authors contribute significantly more than other authors in technical sections, including Results, Experiments, Methods, and Preliminary sections, whereas the last authors disproportionately contribute to conceptual aspects such as Introduction, Discussion, Conception, and Acknowledgments.

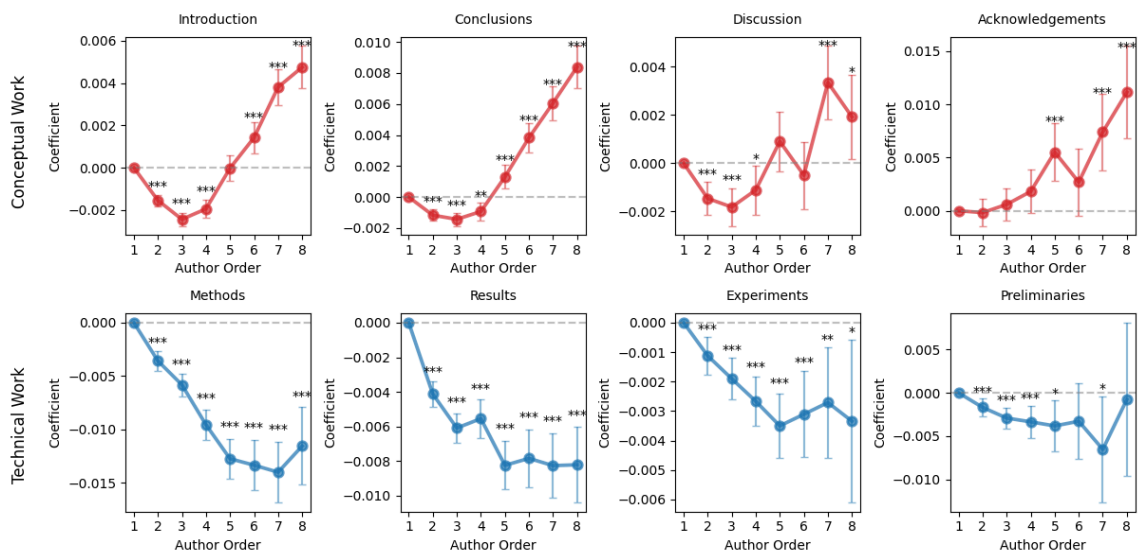

**Figure 1. Hidden Division of Labor in Scientific Teams.** We analyzed 411,808 papers published between 1994 and 2020, identifying sections based on LaTeX \section commands while ignoring \subsection. To assess the relationship between author rank and their focal work represented by the fraction of macros normalized within an author, we estimated regression models within each section. The coefficients represent the relative change in normalized author contributions compared to the first author, with 95% confidence intervals indicated by error bars. All models yielded $P < 0.01$, as denoted by stars in the figure.

## Comparison with Self-Reported Contributions Statements

We validated our dataset against 469 self-reported author contributions from four journals—*Science, Nature, PNAS,* and *PLOS ONE*—confirming its high precision (0.87) and recall (0.71) in identifying paper-writing contributions (Table 1). To compile these data, we first used journal reference information from arXiv to identify papers published in these four journals, which have required author contribution statements for each paper in recent years. We then retrieved each paper's DOI to locate its official homepage on the publisher's website and systematically extracted author contribution statements from acknowledgment sections.

Table 1. Validation of author contributions inferred from LaTeX macros using self-reported data.

| Journal | Sample size N | Time period | Precision | Recall |
|---|---|---|---|---|
| PNAS | 226 | 2007-2023 | 0.93 | 0.76 |
| Nature | 86 | 2010-2023 | 0.68 | 0.66 |
| Science | 12 | 2018-2023 | 0.77 | 0.60 |
| Plos One | 145 | 2008-2023 | 0.91 | 0.66 |
| Total | 469 | 2007-2023 | 0.87 | 0.71 |

## Alignment with First-Author Emphasis

Author order is widely used as a proxy for research contributions[16]. To assess whether contributions align with author order, particularly the conventional assumption that first authors contribute the most, we ran regression models predicting contribution based on author rank while controlling for team size. As shown in Fig. 2, contributions decrease with author order, with the first authors contributing significantly more than all others, forming a declining trend as author order increases. Additionally, the last authors tend to contribute more than the middle authors, reinforcing our earlier observation of their focused work in theoretical sections. This pattern remains consistent across team sizes, thus, further verifying the reliability of our dataset.

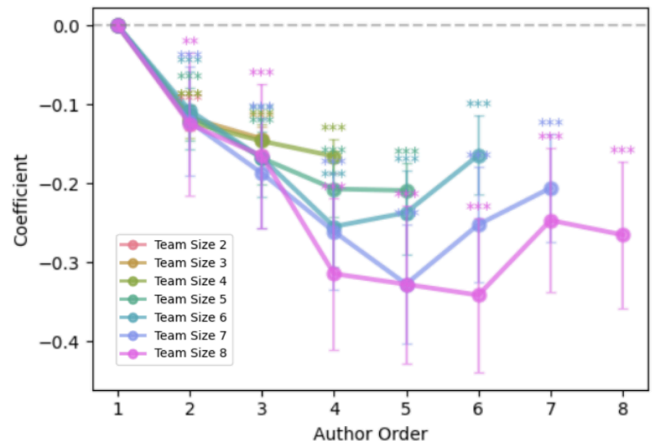

**Figure 2. Declining Contribution with Author Order.** We analyzed 397,521 papers (1994–2020) across different team sizes, covering 96% of all papers. To quantify the relationship between author rank and contribution—measured by the number of unique macros—we ran seven regression models predicting contribution based on author rank within each team size. Coefficients represent the estimated change in contribution relative to the first author, with 95% confidence intervals shown as error bars. All models yielded $P < 0.01$, as indicated by the stars in the figure.

**Alignment with Disciplinary Norms**

Different academic fields follow distinct authorship conventions. In economics, author order is typically alphabetical by last name, while in computer science, the first author is usually the primary contributor, and the last author serves as the senior supervisor[18]. To assess whether writing contributions reflect these disciplinary norms, we ran separate regressions for each field, predicting contribution based on author order. As shown in Fig. 3, in computer science, contributions follow a declining trend, with first authors contributing the most and contributions decreasing with author order. This pattern aligns with the advisor-student model commonly used in the field. In contrast, in economics, contributions remain relatively uniform across author positions, consistent with the alphabetical ordering convention. These findings confirm that our contribution estimates align with established authorship norms, further supporting the reliability of our approach.

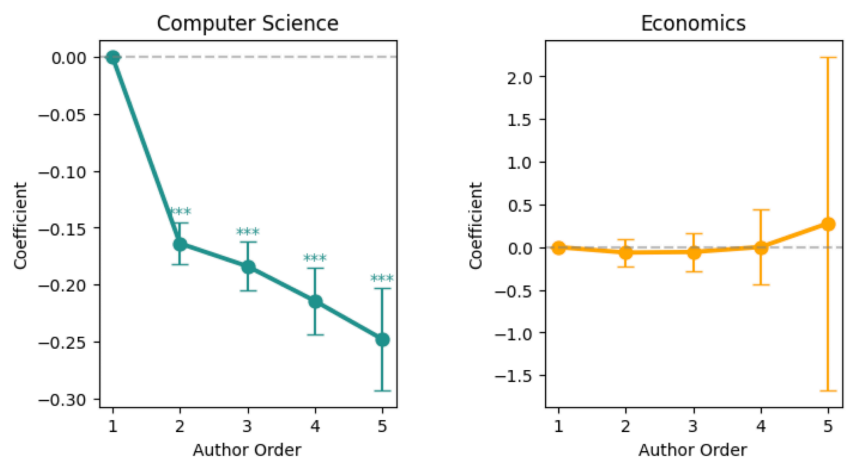

**Figure 4. Discipline-Specific Patterns in Author Order.** We analyzed 76,749 computer science papers (1994–2020) and 1,194 economics papers (1994–2020) based on arXiv domain classifications. To quantify the relationship between author rank and contribution—measured by the number of unique macros—we ran regression models predicting contribution by author rank within each discipline. Coefficients represent the estimated change in contribution relative to the first author, with 95% confidence intervals shown as error bars. All models yielded $P < 0.01$, as indicated by the stars in the figure.

**Validation with Overleaf Editing Data**

To further validate our contribution estimates, we analyzed Overleaf editing histories for two papers, yielding 14 author-paper records. We manually counted each author's edits, ranked them accordingly, and compared these ranks with those derived from our method. A correlation analysis found a Spearman's $\rho = 0.6$ ($p < 0.05$), a meaningful correlation between estimated contributions and actual editing activity, particularly given the small sample size.

**Discussion**

Our study provides the first large-scale evidence of a structured division of labor in scientific teams, showing that writing contributions follow systematic patterns across author order and article sections. By analyzing LaTeX macros, we offer an objective measure of writing contributions, overcoming the limitations of self-reported contribution statements, which are often subjective and inconsistently applied. Our validation confirms the robustness of this approach—high precision and recall compared to self-reported data, strong alignment with disciplinary norms, and a moderate correlation with Overleaf editing histories. These findings support the reliability of our dataset in capturing meaningful authorship patterns.

The structured division of labor we observe advances the understanding of credit allocation in science and has important policy implications. Traditional models assume that contributions decline with author order, but they fail to explain why the last authors often receive more credit, as stated by Merton's *Matthew Effect*, where recognition disproportionately accumulates at the top [1]. Our findings show that while the first authors contribute the most, contributions decline with rank, with a notable increase for the last authors. More importantly, the last authors—especially in large teams—primarily engage in conceptual work. This highlights hidden inequalities in work distribution and credit allocation, raising challenges for recognizing technical contributions by early-career researchers. If junior authors are primarily seen as "muscle" rather than intellectual contributors[9,10], this could hinder their development into independent scholars. As research teams grow larger and temporary scientific positions become more common[16], these trends may further exacerbate career disparities, as recently observed[19].

The limitations of this work point to future directions. Our method estimates contributions based on prior records and assumes that authors consistently adopt and reuse macros in collaborative work. This approach may underestimate the contributions of junior scholars publishing their first papers while potentially overestimating those of senior scholars with extensive publication histories, whether solo or collaborative. Additionally, our method infers general writing contributions from an author's involvement in equation writing, making it most effective in STEM fields with strong analytical components but less applicable in disciplines where equations play a minor role.

These limitations highlight the need to extend this framework beyond author-specific LaTeX macros to behavioral traces of natural language writing, enabling broader coverage across scientific domains. Future research could also analyze actual credit attribution to test the hypothesis that authors performing technical work are less likely to be recognized than their teammates engaged in theoretical work. Despite these limitations, this study provides the first large-scale empirical observation of author contributions and the structured division of labor in scientific teams. As team-based research continues to expand, ensuring fair recognition of contributions will be essential for sustaining the progress of science and innovation.

**Methods**

Established in the 1990s, arXiv.org is the largest preprint repository for STEM fields, including mathematics, statistics, computer science, and more. Since LaTeX is a popular typesetting system for formatting equations in these disciplines, it is widely adopted in arXiv.org submissions. We downloaded 1,600,627 LaTeX source files from papers authored by 2,012,092 unique scientists on arXiv.org between 1991 and 2023. Using this dataset, we identified author contributions by analyzing the LaTeX source code of co-authored papers. If a paper contains a macro previously used by an author, we consider that author a potential contributor. A macro can be attributed to multiple authors if it matches their individual records. We then calculated the number of unique macros contributed by each author and normalized it across all authors on a paper to estimate each author's contribution share. In total, we identified the contributions of 583,817 scientists across 730,914 papers (1991–2023).

Fig. 5 illustrates our approach using a two-author paper and highlights that our methodology relies on two key assumptions The first is the *collaborative learning* assumption: if an author participates in a collaborative project, they are likely to learn and adopt macros used by their coauthors, integrating them into their own future work. This assumption aligns with observed scientific practices, where researchers often reuse LaTeX source code from previous projects. Our analysis also provides empirical support for this assumption. For instance, after collaborating with Huber in 2012, Lindner adopted Huber's macro "\newcommand{\ssm}{\scriptscriptstyle\rm}" as a personalized style for writing subscripts, replacing the standard full form "\scriptscriptstyle\rm," and continued using it in later papers.

The second is the *analytical contribution* assumption: authors who contribute to equations also contribute to general paper writing. This is reasonable because, in research involving analytical work, mathematical content often forms the core of the paper. Authors contributing to this core work are likely to be involved in writing to ensure accurate and clear descriptions of their findings. In the following sections, we verify this assumption by comparing our estimated contributions with self-reported contributions to general writing. We discussed the potential biases introduced by these two working assumptions in the Discussion section.

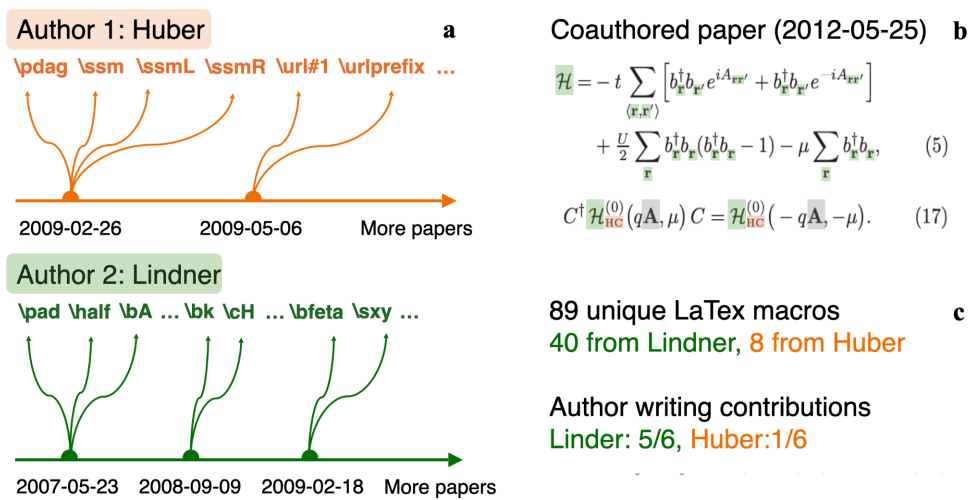

**Figure 5. Estimating author contributions in a coauthored paper based on their history of LaTeX macro usage.** (a) The macro usage history of two scientists, Huber and Lindner. Huber authored 42 papers with a total of 26 unique macros, while Lindner authored 48 papers with a total of 195 unique macros. We construct an individual database of unique macros for each author. (b) We analyze the source code of the 2012 paper coauthored by Huber and Lindner. Using the individual macro databases, we assign macros in the co-authored paper to the respective authors. If a macro appears in both databases, it is assigned to both authors. (c) In the coauthored paper, we identify 8 unique macros from Huber and 40 from Lindner. These numbers are normalized to derive an author writing contribution share of 1/6 for Huber and 5/6 for Lindner.